\documentclass[twocolumn,superscriptaddress,showpacs,aps,floatfix]{revtex4-1}
\usepackage{amsmath,amsfonts,amssymb,epsfig,dcolumn,bm,dsfont,graphics,latexsym,color,graphicx}

\let\oldAA\AA

\renewcommand{\AA}{\text{\normalfont\oldAA}}
\newcommand{\ket}[1]{| {#1} \rangle} 
\newcommand{\bra}[1]{\langle {#1} |} 
\begin{document}
\preprint{AIP/123-QED}
\title{Electrical control of a spin qubit in InSb nanowire quantum dots: strongly suppressed spin relaxation in high magnetic field}
\author{Suzana Miladi\' c}
\affiliation {Faculty of Physics, University of Belgrade, Studentski trg 12, 11001 Belgrade, Serbia}
\author{Pavle Stipsi\'c}
\affiliation {Faculty of Physics, University of Belgrade, Studentski trg 12, 11001 Belgrade, Serbia}
\author{Edib Dobard\v zi\'c}
\affiliation {Faculty of Physics, University of Belgrade, Studentski trg 12, 11001 Belgrade, Serbia}
\author{Marko Milivojevi\'c}
\affiliation {NanoLab, QTP Center, Faculty of Physics, University of Belgrade, Studentski trg 12, 11001 Belgrade, Serbia}
\affiliation{Department of Theoretical Physics and Astrophysics, Faculty of Science,
P. J. \v{S}af\'{a}rik University, Park Angelinum 9, 040 01
Ko\v{s}ice, Slovak Republic}
\begin{abstract}
In this paper, we investigate the impact of gating potential and magnetic field
on phonon induced spin relaxation rate and the speed of the electrically driven single-qubit
operations inside the InSb nanowire spin qubit.
We show that a strong $g$ factor and high magnetic field strength lead to
the prevailing influence of electron-phonon scattering due to deformation potential,
considered irrelevant for materials with a weak $g$ factor, like GaAs or Si/SiGe.
In this regime, we find that spin relaxation between qubit states is significantly suppressed due to the
confinement perpendicular to the nanowire axis.
We also find that maximization of the number of single-qubit operations that can be performed during the lifetime
of the spin qubit requires single quantum dot gating potential.
\end{abstract}
\maketitle
\section{Introduction}
Spin of an electron confined in a semiconductor quantum dot (QD) can act as a carrier of quantum information \cite{BdV}
and a building block of quantum computers.
In order to manipulate electron spin, usage of the external
magnetic~\cite{KBT+06,PLZ+08} and electric~\cite{GBL06,NKN+07,BSO+11} field was suggested.
Although spin control by means
of a magnetic field is straightforward, electrical control of spin qubit through
electric-dipole spin resonance (EDSR) is technologically more desirable~\cite{KGS12,KSW+14,TYO+18,KLS19}.

Spin-orbit coupling (SOC) plays an essential
role in the EDSR spin qubit scheme, since it allows
transitions between qubit states using the spin-independent
driving, such as electric-dipole interaction.
On the other hand, the presence of SOC induces undesired phonon mediated
transitions between qubit states~\cite{KN00,KN01,MSG+02,TFJ02,SL05,PF05,BSF10,LKD+16,S17,S18,LLH+18}.
In order to suppress the coupling to phonons,
approaches like the optimal design of QDs~\cite{CBG+07,MSL16} or the
control of system size~\cite{CM07} was suggested.

Relaxation rates are dependent on the full three-dimensional
QD potential, but in most cases contribution of the
confinement along the direction(s) perpendicular
to the substrate in which QDs are embedded can be neglected.
Assuming magnetic fields up to several tesla,
this reduction is justified in
material with a weak effective Land\'{e} $g$ factor.
A typical example that satisfies
this assumption are lateral GaAs QDs~\cite{GML+08},
while in the opposite direction lies
an InSb nanowire, having two orders of magnitude
stronger $g$ factor~\cite{NCT+09}. Having also very strong SOC,
spin qubits in InSb nanowires~\cite{TGL08,PFB+10,PPB+12,LYS+13,LY14}
have attracted much attention due to the observed~\cite{PFB+10}
fast electric-dipole induced transition between qubit states,
whose speed is equal to the strength of Rabi frequency.

Since both Rabi frequency and phonon induced relaxation rates are dependent
on the magnetic field orientation and strength, design of the gating potential and
SOC, there is a wide range of possibility to
tune their strength, with the goal of obtaining as much as
possible single-qubit operations during its lifetime.

In this paper, we search for the optimal regime
in which electrical control of the InSb
spin qubit can be achieved.
We analyze both single and double quantum dot (DQD) potential
and discuss its positive features and negative drawbacks
on the spin qubit. In the case of double quantum dot potential,
there is the possibility to tune the distance between the dots and
to analyze the effects of the asymmetric gating potential.
Also, we address the situations in which
full three-dimensional confinement has nontrivial influence on
spin relaxation rates.
We will show that scattering by deformation potential dominates in this regime.
Finally, to offer a quantitative insight into the spin qubit quality,
we define a figure of merit as the ratio of Rabi
frequency and the overall spin relaxation rate
and discuss the obtained results in terms of this measure.

This paper is organized as follows.
In Section 2, the single-electron Hamiltonian model of the InSb nanowire is introduced.
In Section 3, we start with the definition of Rabi frequency  and
phonon induced spin relaxation rate between spin qubit states.
After that, we independently study their
dependence
on tunable parameters of the system.
Using the obtained results,
quality of the spin qubit  is
discussed with the help of the figure of merit as a quantitative measure.
In the end, we finish the paper with a short conclusion and the impact of the presented results.
\section{Nanowire spin qubit model}\label{Spin qubit model}
\begin{figure}[t]
\includegraphics[width=8.5cm]{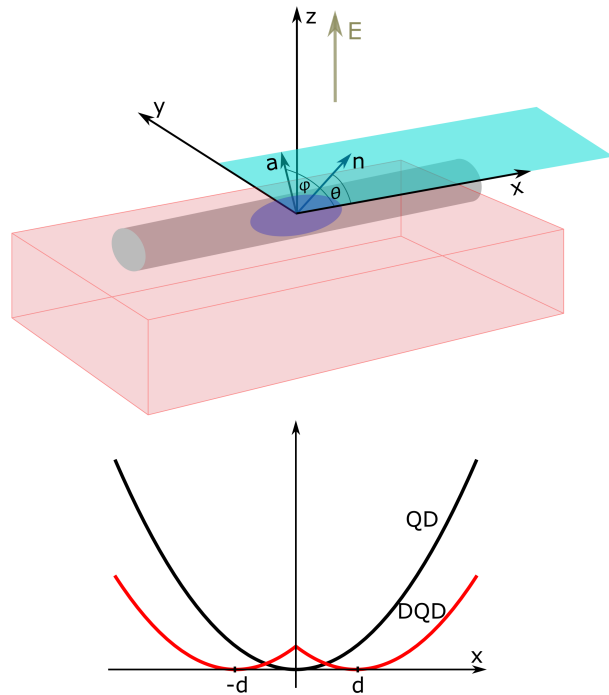}
\caption{\label{NanowireQD}
(upper panel) {\it Nanowire QD - schematic view.} Electron dynamics along the nanowire ($x$) axis
is described by the Hamiltonian $H$, given in Eq.~\eqref{Ham}.
Angle between the nanowire $x$ axis and magnetic field direction ${\bf n}=(\cos{\theta},\sin{\theta},0)$
is equal to $\theta$, while the spin-orbit vector ${\bf a}=(\cos{\varphi},\sin{\varphi},0)$
builds an angle $\varphi$ with the $x$ axis.
(lower panel) Confining potential used in Eq.~\eqref{Ham}:  QD and
DQD potential.
In the case of a DQD potential, Eq. \eqref{DQDho},
symmetric confinement is depicted ($\omega_L=\omega_R$), with
distance between the dots equal to $2d$.}
\end{figure}

We start with the Hamiltonian describing the
electron confined in an InSb nanowire~\cite{LYS+13}
\begin{equation}\label{Ham}
  H=\frac{p^2}{2m^*}+V(x)+H_{\rm so}+H_{\rm z},
\end{equation}
where $m^*$ is the effective mass,
$p=-{\rm i}\hbar\partial/\partial_x$  momentum in $x$ direction,
$V(x)$ is the gating potential used to
localize the electron,
while $H_{\rm so}$ represents the spin-orbit interaction Hamiltonian
consisting of two terms: Dresselhaus~\cite{DR}
and Rashba~\cite{RA}.
The presence of the  Dresselhaus SOC is due to the material
in which an electron is embedded. On the other hand, Rashba SOC appears when an
electric field $E$ in the $z$ direction is applied (see FIG.~\ref{NanowireQD}).
In an InSb nanowire, a spin-orbit interaction Hamiltonian is equal to~\cite{LYS+13}
\begin{eqnarray}\label{SOI}
  H_{\rm so} &=& (\alpha_D\sigma_x+\alpha_R\sigma_y)p,
\end{eqnarray}
where $\sigma_x$ and $\sigma_y$ are Pauli matrices, while $\alpha_D$ and $\alpha_R$ are
Dresselhaus and Rashba spin-orbit coupling strengths.
Suitable change of parameters $\alpha_R$ and $\alpha_D$ with
$\alpha=\sqrt{\alpha_{D}^2+\alpha_{R}^2}$
and $\varphi=\arctan{(\alpha_R/\alpha_D)}$
allows us to write the equation~\eqref{SOI} as
\begin{eqnarray}\label{SOIfin}
  H_{\rm so} &=& \alpha {\bm a}\cdot{\bm \sigma}p,
\end{eqnarray}
using the unit spin-orbit vector ${\bm a}=(\cos{\varphi},\sin{\varphi},0)$ and
the vector ${\bm \sigma}$ made of Pauli matrices.
Finally, $H_{\rm z}$ is the Zeeman term, describing the coupling of spin and
magnetic field
\begin{equation}\label{zeman}
  H_{\rm z}=\frac{g}2\mu_B {\bm B}\cdot{\bm \sigma},
\end{equation}
where $g$ is the effective Land\'{e} factor,
$\mu_B$ is the Bohr magneton,
while ${\bm B}=B{\bm n}$ is the applied magnetic field in the plane of the substrate,
building an angle $\theta$ with the growth $x$ axis of the nanowire
(see the upper panel of FIG.~\ref{NanowireQD}).
In this work a magnetic field is considered to be in-plane to minimize the orbital effects~\cite{MSL16,RSF11,RSF12,WCYW18}.
In Appendix~\ref{1DHam}, we have shown that for $B$ up to 3{\rm T}, orbital effects of a magnetic field are small and can be neglected.

Typical gating that confines a single-electron in experimental setups~\cite{FRR+18} can be modeled as a
harmonic oscillator quantum dot (QD)~\cite{FLK+15} or
double quantum dot (DQD)~\cite{PPB+12} potential.
Corresponding potentials are equal to (see the lower panel of FIG.~\ref{NanowireQD} as an illustration)
\begin{eqnarray}
 V^{\rm QD}(x)&=&\frac{1}2 m^* \omega^2 x^2,\label{QDho}\\
 V^{\rm DQD}(x)&=&\frac{1}2 m^*\;{\rm min}\{\omega_L^2 (x+d)^2,\omega_R^2(x-d)^2\}.\label{DQDho}
\end{eqnarray}

In the case of a QD potential, the only degree of freedom is the harmonic potential
frequency $\omega$, while in the DQD case frequencies $\omega_L$ and $\omega_R$ can be tuned,
as well as the distance $2d$ between the dots.
Since DQD potential allows asymmetric confinement,
we introduce asymmetry parameter $\delta$, equal to the ratio of
frequencies in the left and right dot, $\delta=\omega_{L}/\omega_{R}$.
Impact of the DQD confinement will be discussed in terms of
$\delta$, $2d$, and $\omega_R=\omega$ (more detailed explanation can be found in Section~\ref{EDSR}).

The Hamiltonian of the electron in different potential types and magnetic field strengths
can be solved using the numerical diagonalization~\cite{NS14},
although perturbative approaches in the study of spin qubit properties are common~\cite{TGL08,LYS+13,LLH+18}.
In this work, we follow the numerical approach;
the numerical procedure used in obtaining the
eigenvalues and eigenvectors of the Hamiltonian given in Eq.~\eqref{Ham} is explained
in Appendix~\ref{App}.
In order to successfully diagonalize the Hamiltonian,
orbital $x_0=\sqrt{\hbar/m^*\omega}$ and spin-orbit $x_{\rm so}=\hbar/m^*\alpha$ lengths are defined.
In our calculations, we have used  $m^*=0.014\,m_e$~\cite{PPB+12}, $x_0=30$nm~\cite{PPB+12} and $x_{\rm so}=165$nm~\cite{LWA+18} parameters
for both QD and DQD potentials (recall that $\omega_R=\omega$ in the DQD case),
related to the experimental reports on InSb nanowires.
On the other hand, we have used $g$ factor in bulk InSb material, $g=-51.3$~\cite{CCF88},
being in the range of the experimentally reported values~\cite{FLK+15,NKL+10}.
Initial check of the numerical recipe presented in Appendix~\ref{App}
were exact analytical results obtained in the special case of
the infinite square well~\cite{LL18}.
In this case, we were able to reproduce the results concerning the
angular dependence of the energy splitting between Zeeman sublevels,
Rabi frequency and the relaxation rate.

The Nanowire Hamiltonian, Eq.~\eqref{Ham}, describes the single-electron dynamics in the $x$ direction only.
To ensure the validity of the one-dimensional approximation and to suppress the dynamics
in the $yz$ plane, much stronger $yz$ plane confinement than in the $x$ direction
is needed. In this case, a wave function along both directions, $y$ and $z$,
will correspond to the respective ground state.
To take into the account the wire geometry of the system, the same confinement length $y_0=z_0=10$nm in the $y(z)$ direction is assumed. We model the confinement potential as harmonic~\cite{NS14}, to which the ground state wave function
$\psi(y)={\rm e}^{-y^2/2y_0^2}/\sqrt{\sqrt{{\pi}y_0}}$ corresponds.
In the $z$-direction an additional potential $eEz$ ($z>0$; $z=0$ corresponds to the position of the substrate) is present due to the applied electric field.
Finally, substrate acts as an infinite potential barrier for the confined electron,
forbidding him to propagate in the $z<0$ region~\cite{SS03}.
The ground state, $\psi(z)$, of the Hamiltonian in the $z$ direction
is found using the same numerical method as for the Hamiltonian in the $x$ direction.
Thus, the ground state wave function in the $yz$ plane is equal to $\Psi(y,z)=\psi(y)\psi(z)$.

\section{EDSR and spin relaxation in nanowire spin qubit}\label{EDSR-REL}
In order to achieve electrical control of the nanowire spin qubit,
an oscillating electric field in the $x$ direction
should be switched on, resulting in the Rabi Hamiltonian $H_{\rm R}=e E_0 x \cos(\omega_{E} t)$.
When the applied electric field is in resonance with our quantum system,
Rabi frequency $\Omega_{01}$ is defined as
\begin{equation}\label{Rabi-freq}
  \Omega_{01}=\frac{eE_0}{h}|\bra{0}x\ket{1}|,
\end{equation}
measuring the speed of the single-qubit rotations.
In Eq.~\eqref{Rabi-freq} states $\ket{0}$ and $\ket{1}$ correspond to
the ground and first excited state of the single electron Hamiltonian $H$, while $e|\bra{0}x\ket{1}|$ is the dipole
matrix element.  We are particularly interested in the case where qubit states
are Zeeman sublevels of the orbital ground state, since in this regime
strength of the Rabi frequency can be
manipulated by changing the magnetic field orientation~\cite{LYS+13}.

Besides providing the opportunity to electrically control the spin qubit,
SOC triggers the undesired phonon induced transition between
qubit states, setting up a limit on the qubit lifetime.
Rate of spin relaxation can be determined from the Fermi golden rule
\begin{equation}\label{FGR}
  \Gamma_{01}=\frac{2\pi}{\hbar}\sum_{\nu{\bf q}}|M_{\nu}({\bf q})|^2
  |\bra{\psi_0}{\rm e}^{{\rm i}{\bf q}\cdot{\bf r}}\ket{\psi_1}|^2\delta(\Delta E_{01}-\hbar \omega_{\nu{\bf q}}).
\end{equation}
Transition is triggered by acoustic phonons
of energy $\hbar \omega_{\nu{\bf q}}$ that correspond to the energy
separation between qubit states, $\Delta E_{01}=|E_0-E_1|$.
We assume a linear dispersion relation of acoustic phonons
with respect to the intensity of wave vector $\bf q$, $\omega_{\nu{\bf q}}=c_{\nu}|{\bf q}|$,
yielding $|{\bf q}|=\Delta E_{01}/\hbar c_{\nu}$.

Next, three different geometric factors $|M_{\nu}({\bf q})|^2$ entering
spin relaxation rates originate from different types
of electron-phonon scattering:
electron-longitudinal phonon scattering due to the
deformation potential~\cite{CBG+06}
\begin{equation}\label{la-dp}
|M_{\rm LA-DP}({\bf q})|^2=\frac{\hbar D^2}{2\rho c_{{\rm LA}}V}|{\bf q}|,
\end{equation}
electron-longitudinal phonon scattering due to the piezoelectric field~\cite{CBG+06}
\begin{equation}\label{la-pz}
|M_{\rm LA-PZ}({\bf q})|^2=\frac{32\pi^2\hbar (eh_{14})^2}{\epsilon^2\rho c_{{\rm LA}}V}\frac{(3q_xq_yq_z)^2}{|{\bf q}|^7},
\end{equation}
where $h_{14}$ is piezoelectric constant,
and electron-transverse phonon scattering
due to the piezoelectric field~\cite{CBG+06}
\begin{eqnarray}\label{ta-pz}
|M_{\rm TA-PZ}({\bf q})|^2&=&2\frac{32\pi^2\hbar (eh_{14})^2}{\epsilon^2\rho c_{{\rm TA}}V}\times\\
&&\left|\frac{q_x^2q_y^2+q_x^2q_z^2+q_y^2q_z^2}{|{\bf q}|^5}-\frac{(3q_xq_yq_z)^2}{|{\bf q}|^7}\right|.\nonumber
\end{eqnarray}

Finally, spin relaxation rates are dependent on the transition matrix element $|\bra{\psi_0}{\rm e}^{{\rm i}{\bf q}\cdot{\bf r}}\ket{\psi_1}|^2$
which depends on the full three-dimensional confinement. In order to divide the contribution
of confinements along the nanowire axis and the $yz$ plane, we write the transition matrix element as
$|\bra{0}{\rm e}^{{\rm i}q_x x}\ket{1}|^2|T_{yz}|^2$,
where $|\bra{0}{\rm e}^{{\rm i}q_x x}\ket{1}|^2$ is the contribution along the nanowire direction,
while
\begin{equation}\label{TransitionYZ}
|T_{yz}|^2=\Big|\int\int dy dz |\Psi(y,z)|^2{\rm e}^{{\rm i}(q_y y+q_z z)}\Big|^2
\end{equation}
represents scattering in a plane perpendicular to the nanowire axis.

The role of $|T_{yz}|^2$ in the spin relaxation rate
depends on the regime in which spin qubit operates.
At low magnetic fields, when $|{\bf q}|z_0\ll1$ and
$|{\bf q}|y_0\ll1$, dipole approximation,
${\rm e}^{{\rm i}{\bf q}\cdot{\bf r}}\approx 1+{\rm i}{\bf q}\cdot{\bf r}$,
is valid~\cite{MSL16} and $|T_{yz}|^2$ can be replaced with $(1+|{\bf q}|^2 z_0^2 \cos^2{\theta})\approx1$,
implying that one-dimensional approximation is justified.
However, at higher magnetic fields, dipole approximation is not valid
and confinement in the $yz$ direction can play a significant role.
To determine its role in the spin relaxation rate,
we have numerically calculated $|T_{yz}|^2$ beyond the dipole approximation.

Magnetic field strengths for which the system
operates outside of the dipole approximation ($|{\bf q}|y_0\ge1$) can be roughly estimated;
assuming energy separation between qubit states proportional to $g\mu_B B$,
Fermi golden rule determines phonon wave number
$|{\bf q}|=g\mu_B B/(\hbar c_{\lambda})$,
where $c_{\rm LA}=3800$m/s~\cite{HW86} and
$c_{\rm TA}=1900$m/s~\cite{M87}, giving us magnetic field strengths
for the electron-phonon scattering in the longitudinal (0.084T) and transverse (0.042T) direction above which we are outside of the dipole approximation.

Before we continue, we provide necessary parameters for the
calculation of the spin relaxation rate:
$eh_{14}=1.41\times10^{9}$eV/m~\cite{CBG+06},
$\epsilon=16.5$, $D=7$eV~\cite{DU05},
$\rho=5775{\rm kg}/{\rm m^3}$~\cite{TD}.
\begin{figure}[t]
\centering
\includegraphics[width=8.6cm]{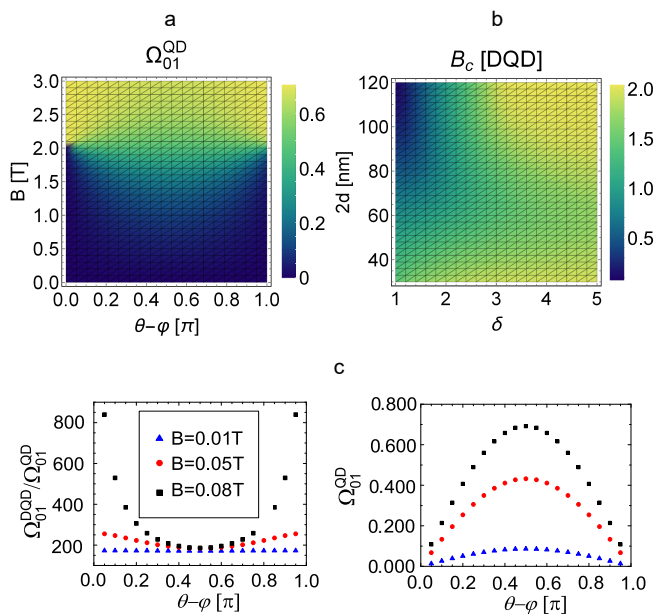}
\caption{\label{Rabi} (a) Dependence of Rabi frequency $\Omega_{01}^{\rm QD}$ (in $e E_0x_0/h$ units)
on $\theta-\varphi\in(0,\pi)$ and $B\in(0,3)$T for QD gating potential.
(b) In the case of DQD confinement, dependence of $B_c$ on the asymmetry parameter $\delta\in(1,5)$
 and distance between the dots $2d\in(30,120)$nm is given.
(c) Dependence of the ratio $\Omega_{01}^{\rm DQD}/\Omega_{01}^{\rm QD}$ on $\theta-\varphi\in(0.05,0.95)\pi$
and magnetic field strengths $B=0.01$T, $B=0.05$T and $B=0.08$T
is presented for the symmetric DQD potential; distance between the dots is equal to $2d=120$nm.
For the same angle range and magnetic field values $\Omega_{01}^{\rm QD}$ in $e E_0x_0/h$ units is presented.}
\end{figure}
\subsection{Rabi frequency}\label{EDSR}
We start the discussion of obtained results with the analysis of
Rabi frequency dependence on the parameters of interest.

In FIG.~\ref{Rabi}a, dependence of $\Omega_{01}$ (in $e E_0x_0/h$ units) on
$\theta-\varphi$
and magnetic field strength is presented for the 
QD confinement potential. Our results confirm the expected $\pi$
periodic behaviour with respect to $\theta-\varphi$~\cite{LYS+13}.
Depending on the magnetic field strength, results can be divided into two classes.
In the first class qubit states represent Zeeman sublevels of the orbital ground state;
in this regime zero Rabi frequency can be found for special magnetic field orientations ($\theta-\varphi=0,\pi$),
since these qubit states have orthogonal spin components.
In the second class, magnetic field strengths have led to rearrangement
of energy levels, such that qubit states originate from the ground and the first excited orbital
state. In this situation, an orbital qubit is constructed, with a very weak dependence
of $\Omega_{01}$ on $\theta-\varphi$ ($\Omega_{01}\neq0$ in the orbital qubit regime for any $\theta-\varphi$).
Critical magnetic field value $B_c$ of spin to orbital qubit transition
is almost independent on $\theta-\varphi$ and can
be easily determined from the eigenspectrum analysis.
Alternatively, for $\theta-\varphi=0,\pi$, abrupt switch of $\Omega_{01}$ from zero to the non-zero value at $B_c$
is a fingerprint of the transition. In the case of the QD potential, we extract
the critical magnetic field value $B_c\approx 2.04$T.

Gating with DQD potential gives qualitatively similar dependence of $\Omega_{01}$ on $B$ and
$\theta-\varphi$.
Being interested in the qualitative comparison
of the impacts of QD and DQD potentials, we first establish a basis for comparison between them.
To this end, we assume the same frequency of the QD potential and the right dot of the DQD potential, $\omega=\omega_R$,
and vary the asymmetry parameter $\delta$ and the distance between the dots $2d$.
For highly asymmetric DQD confinement and the large interdot distance, the electron will
reside on only one dot, i.e. this potential is effectively the same as the single QD potential.
The qualitative similarity of the single and double QD potential is checked
through the comparison of the probability density of the ground and first excited state (qubit states);
similar probability density profiles of the qubit states directly correspond to the
similar Rabi frequency values of the two systems.
Using the numerical comparison of the probability densities and the Rabi frequency
in the case of QD and DQD potential, it can be concluded that for $2d\ge120$nm and $\delta\ge5$
there is no effective difference between the results arising from two potentials.
In other words, one should use $\delta<5$ and $2d<120$nm to test the genuine effects of the
DQD potential.

FIG.~\ref{Rabi}b depicts the dependence of $B_c$ in the DQD case on
$\delta\in(1,5)$ and $2d\in(30,120)$nm.
When compared to the $B_c$ value in the QD case,
drastically lower values are found, especially
in the case of symmetric confinement with well separated
left and right QD. As an example, critical magnetic field value $B_c\approx 0.085$T
for the symmetric DQD confinement with $2d=120$nm is roughly 24 times smaller than in the QD case.

Lower $B_c$ for the symmetric DQD confinement is followed by at most factor 3
increase of $\Omega_{01}(B_c^{\rm DQD})$, when compared to $\Omega_{01}(B_c^{\rm QD})$.
This slight increase, followed by lower $B_c$ below which symmetric DQD operates, indicates
steeper rise of Rabi frequency for symmetric DQD confinements and the possibility
to induce even bigger difference between $\Omega_{01}^{\rm DQD}$ and $\Omega_{01}^{\rm QD}$
for the optimal magnetic field configuration.
To investigate this possibility, we have performed a numerical analysis of
the Rabi frequency ratio $\Omega_{01}^{\rm DQD}/\Omega_{01}^{\rm QD}$ for a wide
range of DQD confinements and different magnetic field strengths/orientations,
such that both systems operate as spin qubits.
Our results confirm that symmetric DQD confinement maximally enhances this ratio
when operating at magnetic field strengths close to $B_c$ for the DQD potential,
while the field orientation should be chosen such
that $\theta-\varphi$ is close to 0 or $\pi$.
In order to illustrate this conclusion, in the left panel of FIG.~\ref{Rabi}c
we present the ratio $\Omega_{01}^{\rm DQD}/\Omega_{01}^{\rm QD}$
for $2d=120$nm and $\delta=1$ in the DQD case, assuming field orientations
$\theta-\varphi\in(0.05,0.95)\pi$ and magnetic field strengths
$B=0.01$T, $B=0.05$T and $B=0.08$T ($B_c^{\rm DQD}\approx 0.085$T for this setup).
Since angles $\theta-\varphi=0,\pi$ should be excluded from the analysis
because they correspond to zero Rabi frequency,
we have restricted our plots to $\theta-\varphi$ region smaller than $\pi$
(see the right panel of FIG.~\ref{Rabi}c for the $\Omega_{01}^{\rm QD}$ values),
obtaining the highest ratio of around $800$. It should be noticed that for
angles closer to $0/\pi$ even bigger ratios ($10^4$) can be obtained,
but at the cost of lowering the value of Rabi frequency.
\subsection{Spin relaxation}
Another important component for determining spin qubit quality
is the spin relaxation rate. Similarly as Rabi frequency, $\Gamma_{01}$ is dependent
on the magnetic field and gating potential.
However, $\Gamma_{01}$ can be additionally dependent on the confinement in $yz$ plane.
In order to compare the influence of three-dimensional confinement
with the confinement along the nanowire axis solely,
we define one-dimensional approximation of the relaxation rate $\Gamma_{01}^{\rm 1D}$
by changing the transition matrix element  $|\bra{\psi_0}{\rm e}^{{\rm i}{\bf q}\cdot{\bf r}}\ket{\psi_1}|^2$
with  $|\bra{0}{\rm e}^{{\rm i}{q_x}x}\ket{1}|^2$ in Eq.~\eqref{FGR}.

It has been known that in lateral GaAs QDs spin relaxation rates are dominated by piezoelectric
field~\cite{CWL04,CSZ+18}. In our case, we wish to analyze
the influence of each relaxation channel;
thus, the overall spin relaxation rate will be divided into three contributions
\begin{equation}\label{reldiv}
  \Gamma_{01}=\Gamma_{01}^{\rm LA-DP}+\Gamma_{01}^{\rm LA-PZ}+\Gamma_{01}^{\rm TA-PZ},
\end{equation}
each dependent on a different geometric factor, see Eqs.~(\ref{la-dp}-\ref{ta-pz}).
\begin{figure}[t]
\centering
\includegraphics[width=4.8cm]{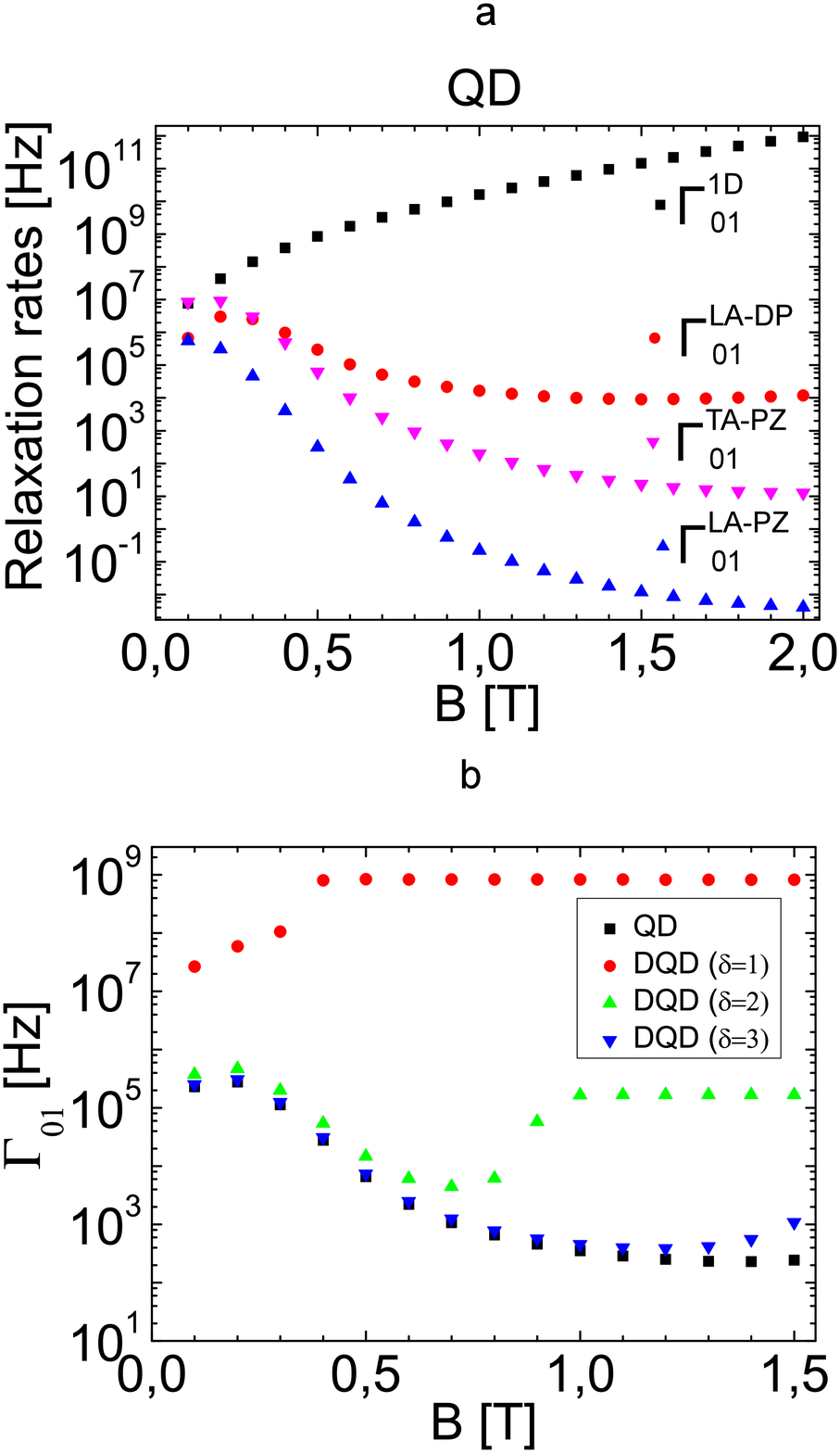}
\caption{\label{Relax} (a) Dependence of the relaxation rates on the magnetic field strength $B\in(0.1,2)$T
for $\theta-\varphi=\pi/2$.
Red circles represent the contribution of deformation potential in the scattering rates,
while inverted pink (blue) triangles show the contribution of piezoelectric field for the electron-phonon scattering in the transverse (longitudinal) direction.
Finally, black squares represent relaxation rates in the one-dimensional
approximation, in which the contribution of the confinement perpendicular to the nanowire axis is neglected.
(b) Dependence of $\Gamma_{01}$ on the magnetic field strength $B\in(0.1,1.5)$T in the case
of QD and DQD confinement potential. Magnetic field orientation is chosen such that $\theta-\varphi=0.05\pi$.
In the DQD case, the distance between the dots is set at $90$nm, while the asymmetry parameter is varied.}
\end{figure}

Before presenting the numerical results, conclusions
independent on the choice of gating potentials are provided.
First, $\Gamma_{01}$ shows oscillatory dependence on the $\theta-\varphi$ angle,
being equal to zero for $\theta-\varphi=0,\pi$ and reaching the maximum for $\theta-\varphi=\pi/2$
in the spin qubit regime~\cite{LLH+18}.
Secondly, for weak magnetic field strengths ($B<0.1$T), piezoelectric field dominate relaxation rates.
At the same time, $yz$-confinement can be ignored.

To explore a new type of behaviour accessible in InSb spin qubits,
we focus our attention on stronger magnetic fields and investigate
its impact on each relaxation channel and one-dimensional approximation of the total relaxation rate $\Gamma_{01}^{\rm 1D}$.
We start from the QD potential.
In FIG.~\ref{Relax}a, dependence of relaxation rates on $B\in(0.1,2)$T
for the fixed angle $\theta-\varphi=\pi/2$ is given~\cite{angles}.
Red circles represent the contribution of deformation potential,
pink inverse (blue) triangles denote the impact of piezoelectric field in the
electron-phonon scattering along the transverse (longitudinal) direction.
Graphs show that relaxation rate $\Gamma_{01}^{\rm LA-PZ}$ can safely be ignored,
while $\Gamma_{01}^{\rm LA-DP}$ and $\Gamma_{01}^{\rm TA-PZ}$ have nontrivial influence on $\Gamma_{01}$.
For weak magnetic fields $\Gamma_{01}^{\rm TA-PZ}$ term is dominant, while
for large magnetic fields $\Gamma_{01}^{\rm LA-DP}$ should be considered solely~\cite{NS14}.
Different influence of $\Gamma_{01}^{\rm TA-PZ}$ and $\Gamma_{01}^{\rm LA-DP}$
lies in the opposite behaviour of the corresponding geometric factors:
$|M_{\rm TA-PZ}({\bf q})|^2$ ($|M_{\rm LA-DP}({\bf q})|^2$) is inversely (directly) proportional to the energy splitting between
the Zeeman levels and decreases (increases) with the magnetic field rise.

Contribution of the $yz$ plane confinement on the spin relaxation rate can be determined by comparing the
$\Gamma_{01}^{\rm 1D}$ with relaxation rate channels. The comparison is illustrated in FIG.~\ref{Relax}a,
clearly demonstrating that one-dimensional approximation of the spin relaxation rate is valid
only for weak magnetic fields, below 0.1T. At higher fields, due to the strong $g$ factor of the
InSb material, both $|{\bf q}|y_0$ and $|{\bf q}|z_0$ are greater than one, triggering the effects of the $yz$ plane
confinement for each relaxation rate channel. Thus, suppressed spin relaxation represents a
fingerprint of a material with a strong g factor.

In the case of DQD potentials, dependence of $B_c$ on the form of gating
presents a serious limitation on the regimes that can be accessed.
For example, if the $B_c$ value is sufficiently weak,
$B_c<0.1$T, the spin qubit operates
under the dominant influence of the piezoelectric field.
A strong magnetic field regime is beneficial for spin qubit operation
due to strong Rabi frequency and suppressed spin relaxation.
In order to operate in this regime, asymmetric DQD potential should be used.
To compare the influence of QD and DQD potential on $\Gamma_{01}$,
in FIG.~\ref{Relax}b, we plot the dependence of
the spin relaxation rate in the case of QD and DQD confinement on the
magnetic field strength $B\in(0.1,1.5)$T, assuming $\theta-\varphi=\pi/2$ and $2d=90$nm.
Besides the symmetric $\delta=1$ confinement, asymmetric DQD confinements
$(\delta=2,3)$ were analyzed as well. The presented results
show that DQD gating leads to increased relaxation rates, when compared to the QD potential.
This difference is minimized for highly asymmetric gating potentials.
Note that $B$ independent $\Gamma_{01}$ values
suggest that orbital qubit is created: energy difference between the states
with the same spin component (representing the orbital qubit states in our case)
is independent on $B$ and triggers phonons on the same energy,
leading to the observed effect.
Consequently, these points should be excluded from the spin qubit analysis.

Finally, we emphasize that in the special case of the asymmetric DQD potential with $\delta=1.5$
similar trend of the spin relaxation rate is ascertained~\cite{LLH+18}, i.e. after the increase of the spin relaxation
rate in the dominant regime of the piezoelectric field, suppression of spin relaxation is observed, followed by the
increase up to magnetic field independent saturation value (see the green triangles in FIG.~\ref{Relax}b as a comparison).

%
\begin{figure}[t]
\centering
\includegraphics[width=8.5cm]{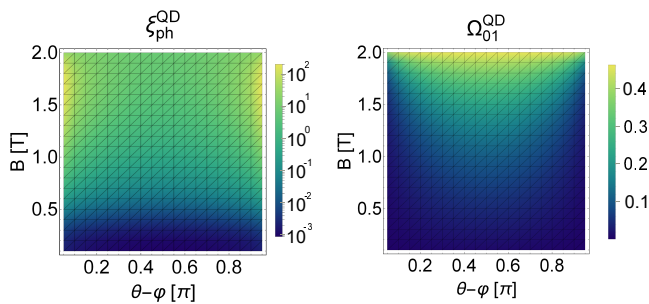}
\caption{\label{merit} For the QD confining potential, dependence of the figure of merit
$\xi_{\rm ph}^{\rm QD}$ (given in dimensionless unit 7.25 ${\frac{\rm m}{\rm V}}\times E_0$)
and Rabi frequency (in $e E_0x_0/h$ units)
on the relative angle $\theta-\varphi\in(0.05,0.95)\pi$
and magnetic field strength $B\in(0.1,2)$T is presented.}
\end{figure}
\subsection{Spin qubit quality}
Quantitative estimate of the spin qubit quality can be given with the help of
the figure of merit $\xi$~\cite{MSL16}
\begin{equation}\label{Merit1}
  \xi=\frac{\Omega_{01}}{\Gamma_{01}+\Gamma_{\rm o}},
\end{equation}
measuring the number of qubit operations that can be implemented during the qubit lifetime.
In Eq.~\eqref{Merit1} $\Gamma_{\rm o}$ represents relaxation rate of
decay channels different from phonons.
To divide the contribution of
phonons from them, we rewrite $\xi$ in terms of the
phonon figure of merit $\xi_{\rm ph}=\Omega_{01}/\Gamma_{01}$
and relative influence of other channels with respect to phonons $\Gamma_{\rm o}/\Gamma_{01}$.
Thus,
\begin{equation}\label{Merit2}
  \xi=\frac{\xi_{\rm ph}}{1+\frac{\Gamma_{\rm o}}{\Gamma_{01}}}.
\end{equation}

We first analyze $\xi_{\rm ph}$ for the QD confinement.
Neglecting the weak magnetic field regime~\cite{uglovi},
in FIG.~\ref{merit} we present
the dependence of $\xi_{\rm ph}^{\rm QD}$ on
$B\in(0.1,2)$T and $\theta-\varphi\in(0.05,0.95)\pi$.
Restricted $\theta-\varphi$ domain plotted
is due to the a priori exclusion of $\theta-\varphi=0,\pi$ values
($\Gamma_{01}^{\rm QD}=0$ in these situations).
Plots show that to maximal value of $\xi_{\rm ph}^{\rm QD}$
correspond relative angles $\theta-\varphi=0.05\pi, 0.95\pi$.
This result suggests that for $\theta-\varphi$ closer to $0$ or $\pi$ than presented
even bigger $\xi_{\rm QD}$ values can be obtained, at the cost of lowering the Rabi frequency.
In other words, $\Gamma_{01}^{\rm QD}$ has a steeper decline to zero
than $\Omega_{01}^{\rm QD}$, when $\theta-\varphi$ goes from $\pi/2$ to $0$ or $\pi$.

Magnetic field orientation isotropy of $\Gamma_{\rm o}$~\cite{CSZ+18}
implies that shift from $\theta-\varphi=\pi/2$ increases $\Gamma_{\rm o}/\Gamma_{01}^{\rm QD}$ also.
Thus, in order to maximize $\xi$, optimization of both $\xi_{\rm ph}^{\rm QD}$ and $\Gamma_{\rm o}/\Gamma_{01}^{\rm QD}$ is needed.
Since at high magnetic fields phonon induced relaxation dominates~\cite{CSZ+18},
deviation of $\theta-\varphi$ from $\pi/2$ improves the spin qubit quality until
$\Gamma_{\rm o}/\Gamma_{01}^{\rm QD}$ drops below 1.
This sets up the optimal magnetic field orientation.

Finally, we compare the impacts of DQD and QD potentials on the spin qubit quality.
As discussed in Section~\ref{EDSR}, Rabi frequency
in the DQD case can be three orders of magnitude
greater than in the QD case. Enhanced Rabi frequency suggests that SOC
effects are more pronounced; thus, phonon induced spin relaxation rate should be enhanced.
When compared to the QD case, an increase of $\Gamma_{01}^{\rm DQD}$ followed by the negative
trend of $\xi_{\rm ph}^{\rm DQD}$ ensures that spin qubit quality decreases;
symmetric DQD confinements give the poorest results,
while highly asymmetric DQD potentials provide similar values as for QD gating.
\section{Conclusions}
We have investigated the influence of gating potentials,
magnetic field strength and orientation on Rabi frequency
and spin relaxation rate
in a single electron InSb nanowire spin qubit.
Due to the strong Land\'{e} $g$ factor, we were able to show that
InSb spin qubit can operate in the
regime in which deformation potential of acoustic phonons dominate relaxation rate.
Qualitatively new behavior of spin relaxation rate
comes from the confinement perpendicular to the nanowire axis,
offering a new regime in which spin qubit can successfully operate.
We have shown that gating potential has a crucial role in enabling such a situation,
additionally pointing out simple harmonic potential as
beneficial for the optimal definition of a spin qubit.
Although presented for InSb nanowire spin qubits, conclusions remain
valid for spin qubits in other materials with a strong $g$ factor.
Thus, modifications of $g$ due to different effects,
e.g. strong in-plane magnetic field~\cite{SHS+18},
do not interfere with the conclusions stated in this work.
\acknowledgments
This research was funded by the Ministry of Education, Science, and Technological Development of the Republic of Serbia under projects
ON171035 and ON171027 and the National Scholarship Programme of the Slovak Republic (ID 28226).
\appendix
\section{Derivation of the effective one-dimensional Hamiltonian}\label{1DHam}
Here, we derive the effective one-dimensional Hamiltonian $H$ of the electron in an InSb quantum wire,
by averaging the three-dimensional kinetic energy term $T_{\rm 3D}$ and two-dimensional spin-orbit Hamiltonian $H_{\rm so}^{\rm 2D}$
over $y$ and $z$ direction. Thus, we start from the three-dimensional Hamiltonian
\begin{equation}\label{3DHam}
  H_{\rm 3D}=T_{\rm 3D}+V(x)+H_{\rm so}^{\rm 2D}+H_{\rm z},
\end{equation}
where $T_{\rm 3D}=\sum_{i=x,y,z}P_i^2/2m^*$ ($P_i=p_i+eA_i$),
\begin{eqnarray}
H_{\rm so}^{\rm 2D}&=&\alpha_R\Big(P_x\sigma_y-P_y\sigma_x\Big)+\alpha_D\Big(P_x\sigma_x-P_y\sigma_y\Big),\,\,\,
\end{eqnarray}
while $V(x)$ and $H_{\rm z}$ are the gating potential and the Zeeman term, defined in Eq.~\eqref{zeman} and Eqs.~(\ref{QDho}-\ref{DQDho}), respectively.
The choice of the vector potential components $A_x=-Bz\sin\theta$, $A_y=0$, $A_z=-By\cos{\theta}$ is such that
corresponds to the applied in-plane magnetic field ${\bf B}=B(\cos{\theta},\sin{\theta},0)$.
After averaging the kinetic energy operator over the $y$ and $z$ direction
using the ground state wave function $\Psi(y,z)=\psi(y)\psi(z)$, we get
\begin{eqnarray}
  {\langle T\rangle}&=&\frac{p_x^2}{2m^*}-\frac{eB {\langle z\rangle}\sin\theta}{m^*}p_x+\Big[\frac{\langle p_y^2\rangle}{2m^*}+\nonumber\\
&&\frac{\langle( p_z -eBy\cos{\theta})^2\rangle}{2m^*}
 +\frac{e^2B^2\sin^2{\theta}\langle z^2\rangle}{2m^*}\Big].\label{effKIN}
\end{eqnarray}
In the previous equation, only the first and second term affect the dynamics in the $x$ direction,
while all terms in the square brackets can be considered the constant shift of energy and,
therefore, can be neglected.

Next, effective one-dimensional spin-orbit interaction Hamiltonian is equal to
\begin{eqnarray}
  {\langle H_{\rm so}\rangle}&=&\alpha_R\Big((p_x-eB\langle z \rangle \sin\theta)\sigma_y-\langle p_y \rangle \sigma_x\Big)\nonumber\\
&&+\alpha_D\Big((p_x-eB\langle z \rangle \sin\theta)\sigma_x-\langle p_y \rangle\sigma_y\Big)\nonumber\\
&=&(p_x-eB\langle z \rangle \sin\theta)(\alpha_R\sigma_y+\alpha_D\sigma_x)\label{effSOC},
\end{eqnarray}
where we have used the fact that expectation value of the momentum $p_y$,
$\langle p_y \rangle=\int_{-\infty}^{\infty}\Psi^{*}(y,z)p_y\Psi(y,z)$, is explicitly equal to zero.

A further simplification of the effective Hamiltonian can be made by neglecting the term $eB {\langle z\rangle}\sin{\theta}p_x/m^*$
from the Eq.~\eqref{effKIN} and $eB\langle z \rangle \sin\theta$ from Eq.~\eqref{effSOC}.
Assuming that intensity of $p_x$ is proportional to $\hbar/x_0$, magnetic field dependent terms can be neglected if the
relation
\begin{equation}
\frac{\hbar}{x_0}\gg eB {\langle z\rangle}
\end{equation}
is satisfied. More concretely,
when the $\hbar/x_0$ is for a factor of 10 stronger than the
magnetic field dependent term, orbital effects of the magnetic field are small and can be discarded.
In our calculations, the magnetic field strengths of interest are
up to 3T, yielding the relation for the $z$ expectation value
\begin{equation}
{\langle z\rangle}\leq0.1\frac{\hbar}{e x_0\times 3{\rm T}}
\end{equation}
that has to be satisfied to successfully operate in this regime.
As discussed in Section~\ref{Spin qubit model},
the wave function $\psi(z)$ is dependent on the strength of the applied electric field $E$:
with the increase of the electric field strength $\langle z\rangle$ increases. In other words, the strength of the electric field is limited from above.
Numerical estimate for the critical value of electric field is $6.5\times10^6\, {\rm V}/{\rm m}$, going to be used in our numerical calculations.
Under these assumptions, the effective one-dimensional Hamiltonian resembles the one defined in Eq.~\eqref{Ham}, used in the rest of the paper.
\section{Numerical solution of the one-dimensional Schr\"{o}dinger equation}\label{App}
In order to find eigenvectors and eigenenergies of the
Hamiltonian $H$, given in Eq.~\eqref{Ham}, numerical
diagonalization is performed.
After defining orbital and spin-orbit lengths as $x_0$ and
$x_{\rm so}=\hbar/m\alpha$, respectively,
such that $x=x_0u$, where $u$ is dimensionless variable,
$H$ can be written in the following form
\begin{equation}\label{red}
  H=\frac{\hbar^2}{2m^*x_0^2}H_{\rm red}.
\end{equation}
Eigenvectors of $H$ are the same as of
$H_{\rm red}$, while eigenvalues of $H$ and $H_{\rm red}$
differ for the factor $\hbar^2/2m^*x_0^2$, having the energy units.
The benefits of using $H_{\rm red}$ instead of $H$ stems from the transfer
into dimensionless units, more suitable for numerical manipulation.
The concrete form of $H_{\rm red}$ is equal to
\begin{equation}
H_{\rm red}=-\frac{d^2}{du^2}-2{\rm i}\frac{x_0}{x_{\rm so}}{\bf a}\cdot{\bm \sigma}\frac{d}{du}
+V_{\rm eff}(u)+g_{\rm eff}{\bm n}\cdot{\bm \sigma},
\end{equation}
where $g_{\rm eff}$ and $V_{\rm eff}(u)$ are effective Land\'{e} factor and effective potential,
respectively,
\begin{equation}
g_{\rm eff}=g\frac{m^*x_0^2\mu_BB}{\hbar^2},\;\;V_{\rm eff}(u)=\frac{2m^*x_{0}^2}{\hbar^2}V(x_0u),
\end{equation}
while vectors ${\bf a}$ and ${\bf n}$ are spin-orbit and magnetic field unit vectors, respectively,
defined in the main text.
The form of effective potential depends on the choice of gating potential~(\ref{QDho}-\ref{DQDho}),
while effective Land\'{e} factor is linearly dependent on the magnetic field strength $B$.

To numerically solve eigenproblem of $H_{\rm red}$,
orbital space is discretized with an uniform grid.
First and second derivative of a wave function are approximated
by finite difference uniform grid formulas~\cite{NUM}
\begin{eqnarray}
  \frac{d\psi(u)}{du} &=& \frac{\psi_{-4}}{280h}-\frac{4\psi_{-3}}{105h}+\frac{\psi_{-2}}{5h}-\frac{4\psi_{-1}}{5h}\nonumber\\
  &-&\frac{\psi_{4}}{280h}+\frac{4\psi_{3}}{105h}-\frac{\psi_{2}}{5h}+\frac{4\psi_{1}}{5h}+O(h^8),\label{prvaJNA}
\end{eqnarray}
\begin{eqnarray}
  \frac{d^2\psi(u)}{du^2}&=&-\frac{\psi_{-4}}{560h^2}+\frac{8\psi_{-3}}{315h^2}-\frac{\psi_{-2}}{5h^2}+\frac{8\psi_{-1}}{5h^2}-\frac{205\psi_{0}}{72h^2}\nonumber\\
  &-&\frac{\psi_{4}}{560h^2}+\frac{8\psi_{3}}{315h^2}-\frac{\psi_{2}}{5h^2}+\frac{8\psi_{1}}{5h^2}+O(h^8),\label{drugaJNA}
  \end{eqnarray}
with accuracy to the $h^8$ order, where $h$ is the uniform grid step.
By definition, $\psi_{\pm n}=\psi(u\pm nh)$ represent wave functions shifted in the
left/right (-/+) direction of the coordinate space for $nh$.

Uniform grid formulas allow us to represent Hamiltonian as a square matrix.
Effective potential is represented as a diagonal matrix,
while matrix representation of the first and second order derivative
have nondiagonal terms in addition.
Since $H_{\rm red}$ is dependent on spin degrees of freedom also,
orbital part of the Hamiltonian is trivially extended in the spin space.
Also, Zeeman Hamiltonian is trivially extended in the orbital space,
while matrix form of spin-orbit Hamiltonian is obtained
as a tensor product of the first derivative matrix and spin
Hamiltonian ${\bf a}\cdot{\bm \sigma}$.

In the QD case, harmonic potential is centered at $u=0$,
while in the case of DQD potential numerical calculations assumed
each QD center range from $u=\pm1/2$ to $u=\pm2$.
We have checked that for all studied situations
choice of $u$ from the interval $(-8,8)$
is enough to capture the smooth decline of the orbital wave function
to 0 at $u=\pm8$.
Also, the division of the orbital space into $N=2000$ parts
was enough to ensure convergence of the results, i.e.
for the increase of $N$ to 4000 relative difference between the results is below $10^{-4}$.


\begin{thebibliography}{50}
\bibitem{angles} Although dependence of relaxation rates on the magnetic field strength are presented
                 for the special $\theta-\varphi=\pi/2$ magnetic field orientation, conclusion remain
                 valid for different $\theta-\varphi$ angles.
\bibitem{uglovi} Low fields correspond to weak Zeeman splitting. In this case
                 low temperatures are needed, such that Zeeman splitting
                 is above the thermal broadening; temperature dependence
                 analysis is beyond the scope of the presented work.
\end{thebibliography}
\end{document}